\documentclass[final,5p]{elsarticle}
\usepackage{jasr}
\usepackage{framed,multirow}
\usepackage{amssymb}
\usepackage{natbib}
\usepackage{latexsym}
\usepackage{url}
\usepackage{amsmath}
\usepackage{subcaption}
\usepackage{xcolor}
\definecolor{newcolor}{rgb}{.8,.349,.1}
\usepackage[citebordercolor=white]{hyperref}

\journal{Advances in Space Research}

\begin{document}

\begin{frontmatter}

\title{Properties of 2017-18 `failed' Outburst of GX 339-4} 

\author[1]{Dipak \snm{Debnath}\corref{cor1}}
\cortext[cor1]{Corresponding author: 
  email: dipakcsp@gmail.com}
\author[1]{Kaushik \snm{Chatterjee}}
\ead{mails.kc.physics@gmail.com}
\author[1]{Sujoy Kumar \snm{Nath}}
\author[2]{Hsiang-Kuang \snm{Chang}}
\author[1]{Riya \snm{Bhowmick}}

\address[1]{Indian Centre For Space Physics, 43 Chalantika, Garia Station Road, Kolkata, 700084, India}
\address[2]{Institute of Astronomy, National Tsing Hua University, Hsinchu 300044, Taiwan}


\begin{abstract}

The Galactic transient black hole candidate GX 339-4 is a very interesting object to study as it showed both complete and failed 
types of outbursts. We studied both spectral and temporal properties of the 2017-18 outburst of the source using archival data of 
NICER and AstroSat instruments. This 2017-18 outburst is found to be failed in nature, as during the entire period of the outburst, 
the source was only in the hard spectral state. Source spectra were highly dominated with the non-thermal fluxes. When we tried to 
fit spectra with phenomenological models, most of the spectra were fitted with only the powerlaw model, and only six spectra required 
disk black body plus powerlaw models. While fitting spectra with the physical two-component advective flow (TCAF) model, we observed 
that the flow was highly dominated by the sub-Keplerian halo rate. The presence of stronger shock at a larger radius from the black 
hole was also observed during the rising and declining phases of the outburst. A prominent signature of $0.31$~Hz QPO is observed with 
its four harmonics. Mass of the black hole was also estimated from our spectral analysis with the TCAF model as $10.76^{+0.77}_{-1.07}~M_\odot$.

\end{abstract}


\end{frontmatter}

\section{Introduction}

Galactic stellar massive transient black hole X-ray binaries (BHXRBs) are very interesting objects to study as they show rapid evolution 
in their spectral and temporal properties. A strong correlation between these two properties is seen in the BHXRBs. In the temporal 
domain, they show an evolution of their hard and soft band X-ray counts/fluxes. Sometimes they show time and phase lags between hard and 
soft band photons. Observation of low and high (occasionally found) frequency quasi-periodic oscillations (QPOs) are one of the important 
temporal features. They are found in the power-density spectrum (PDS), obtained from the fast-Fourier transformation of the light curves. 
In an outburst, we generally see four spectral states, such as hard (HS), hard-intermediate (HIMS), soft-intermediate (SIMS), soft (SS), 
which forms hysteresis loop in the following sequence : HS $\rightarrow$ HIMS $\rightarrow$ SIMS $\rightarrow$ SS $\rightarrow$ SIMS 
$\rightarrow$ HIMS $\rightarrow$ HS. But in a failed or harder type of outburst, soft states (sometimes even intermediate states) are missing. 
Outflow/jet is also a common feature of these black hole candidates (BHCs). The nature of the QPOs, jets are found to vary with the spectral 
states (see, Remillard \& McClintock 2006; Debnath 2018 for reviews). Type-C QPOs are commonly observed in the harder (HS and HIMS) spectral 
states, whereas type-B or A QPOs are found in the SIMS. Low-frequency QPOs are non-observable in the SS. Furthermore, sometimes in the rising 
and declining harder spectral states monotonic evolution of the type-C QPOs are observed during an outburst of a transient BHC. In the SIMS, 
QPOs are generally observed sporadically. In a jet dominated source, compact jets are observed in the harder spectral states, 
predominately in the hard state. Discrete or blobby jets are observed in the intermediate spectral states. 

Although there is a conflict about the triggering of an outburst, according to Ebisawa et al. (1996) an outburst in a transient BHC is 
triggered by the sudden increase of viscosity at the outer edge of the accretion disk. When viscosity becomes weaker, we see that BHCs 
move toward the declining phase of their outbursts. Recently, Chakrabarti et al. (2019), Bhowmick et al. (2021), and Chatterjee et al. (2022) 
discussed a possible relation between the quiescence and outburst phases of three well known recurring transient BHCs H~1743-322, 
GX~339-4 and 4U 1630-472 respectively. According to them, matter supplied by the companions starts to pile up at a pile-up radii ($X_p$) located 
at the outer disk during the quiescence phases of these transient BHCs. An outburst could be triggered by a sudden rise in viscosity at the 
temporary reservoir at $X_p$. A good correlation between outburst and quiescence periods are found for these sources. The nature (duration, 
peak numbers, highest peak flux, etc.) of the outbursts are now more clearly understood. In a failed outburst, it is often found that the matter 
accumulated at $X_p$ before the outburst is not cleared completely. Fresh matter is added to the leftover matter to trigger the next failed 
or complete outburst.

To understand accretion flow dynamics around a black hole, one needs to study both spectral and temporal properties with a physical model. 
Recent studies of our group have found that the two-component advective flow (TCAF) model is quite successful to explain physics around these 
compact objects. Accretion flow dynamics of these objects are now more clear from the variation of the model fitted physical flow parameters 
(Keplerian disk rate $\dot{m_d}$, sub-Keplerian halo rate $\dot{m_h}$, location of the shock $X_s$, and compression ratio $R$) of more than 
fifteen BHCs (see, Debnath et al. 2014, 2015a,b, 2017; Chatterjee et al. 2016, 2019, 2020, 2021a,b; Jana et al. 2016, 2017; Shang et al. 2019). 
Origin of the QPOs are well understood from the TCAF model fitted shock parameters (Debnath et al. 2014; Mondal et al. 2015; Chakrabarti et al. 2015; 
Chatterjee et al. 2016, 2021b). One can also estimate black hole mass and contribution of jet component of the X-ray fluxes into the total observed 
X-ray, etc. (Debnath \& Chatterjee et al. 2021; Jana et al. 2020, 2021; Nath et al. 2021) from the spectral analysis with the model. 2010, 2016 
outbursts of this source were also studied with the TCAF model in past (Debnath et al. 2015a; Mondal et al. 2016).

Galactic transient BHC GX 339-4 was discovered in 1973 by the MIT X-ray detector onboard OSO-7 satellite (Markert et al. 1973). This low mass 
BHXRB system has a BH mass function of $M_{bh}~sin(i)$ = $5.8\pm0.5~M_\odot$ with companion mass $m$ = $0.52~M_\odot$. It is located at a 
distance of $d~\geq~6$~kpc with R.A.=$17^h02^m49^s.56$ and Dec.=$-48^\circ46'59''.88$ (Hynes et al., 2003, 2004). Parker et al. (2016) has
estimated mass and distance of the source to be $9.0^{+1.6}_{-1.2}$~$M_\odot$ and $8.4\pm0.9$~kpc from the spectral analysis.
Sreehari et al. (2019) also estimated the mass of the source as $8.28-11.89$~$M_\odot$ from the spectro-temporal analysis of its multiple outbursts.
In the last two and half decades, it showed five major complete/normal outbursts during the RXTE era (1996-2011), whereas in the MAXI era until 2020 
(2009-2020), it showed three complete and three failed outbursts (for more details see, Bhowmick et al. 2021). This source has been extensively 
studied in multiwavelength bands during the complete outbursts. 

This outburst was first reported by Russell, Lewis \& Gandhi (2017) in the optical band using Australia's 2m Faulkes Telescope South. 
Subsequently, this outburst was monitored in X-ray (Gandhi et al. 2017; Remillard et al. 2017), radio (Russell et al. 2017) and other wavebands.
Garcia et al. (2019) also examined X-ray reflection near the black hole with NuSTAR and Swift data of the 2017 failed outburst of GX 339-4.
Wang et al. (2020) analyzed the NICER and NuSTAR spectra during the 2017 and 2019 outbursts using relativistic and distant reflection models. 
They also observed a reverberation lag of $9\pm3$ ms. Husain et al. (2021) also analyzed the 2017 and 2019 outbursts of this source using 
AstroSat data. Recently, de Haas et al. (2021) presented quasi-simultaneous radio and X-ray observations for 2017--18 outburst. 
They established a correlation between the radio and X-ray fluxes during the outburst, which suggested a flatter correlation between them. 
This indicates an inefficient coupling between the outflow and inflow that supports the existence of hard states only. 
Here, we have re-looked at the 2017--18 failed outburst of the source. To get a more clear picture of the accretion flow dynamics during 
the outburst, we studied it using the TCAF model.

The \textit{paper} is organized in the following way. In the next Section, observation and data analysis procedures are discussed. 
In \S 3, the results obtained from our analysis, first temporal and then spectral are discussed. In the \S 4, a discussion and conclusion 
based on our analysis are presented. Finally, a summary of our results is discussed in the \S 5.

\section{Observation and Data Analysis}

We studied the $2017-18$ outburst of GX 339-4 using archival data of MAXI/GSC, Swift/BAT, NICER/XTI, and AstroSat/LAXPC instruments. 
To study the nature of the outburst profile in soft and hard X-ray bands and evolution of the hardness ratio, we used MAXI/GSC (in $2-10$~keV band) 
and Swift/BAT (in $15-50$~keV band) during a period of $\sim 7$~months from 2017 August 25 (MJD=57990) to 2018 March 20 (MJD=58197). 
For the detailed spectral and temporal analysis, we used both NICER and AstroSat data. 

NICER monitored the $2017-18$ outburst of the BHC GX 339-4. However, observations are not present on a daily basis. There is a data gap
of close to 2.5 months from 2017 November 3 (MJD 58060) to 2018 January 18 (MJD 58136). We have checked for archival Swift/XRT data also. 
However, we have found a similar period of data gap there also. Depending on the noise level of spectra, we have used a total of 21 observations 
from NICER and 2 observations (orbits) from AstroSat/LAXPC data. For this work we have used $2-10$~keV MAXI/GSC, $15-50$ keV Swift/BAT, 
$0.6-10$ keV NICER data. For broadband analysis, we have also used archival AstroSat/LAXPC data. The data reduction process for different 
satellites/instruments is given below.

\subsection{NICER}

NICER has unprecedented spectral and timing resolutions of $\sim 85$ eV at 1 keV and $\sim 100$ nanoseconds respectively. We have used the 
latest calibration files (20210707) to run the data reduction process. First, we run the `\textit{nicerl2}' script which runs a standard 
pipeline that produces level-2 cleaned event files using standard calibration. Then we run the `\textit{barycorr}' command which applies 
barycenter correction to the cleaned event files. Using the barycenter corrected event files, we extract light curves of desired binsize 
using {\fontfamily{qcr}\selectfont XSELECT} task. We also extract spectrum files from {\fontfamily{qcr}\selectfont XSELECT} using the 
cleaned event file. At last, we run the \textit{nibackgen3c50} command which produces background spectra corresponding to each observation ids. 

To search for QPOs, we initially extracted 1s binned lightcurves from the barycenter corrected event files. There were large data gaps in the 
resulting lightcurves. In order to select good time intervals (GTIs) with continuous data, we used time filtering in 
{\fontfamily{qcr}\selectfont XSELECT}. We extracted 0.01s binned lightcurves from the selected GTIs. The {\fontfamily{qcr}\selectfont powspec} 
task of {\fontfamily{qcr}\selectfont XRONOS} software package was then used to generate the white noise subtracted power density spectra (PDS).
Each lightcurve was divided into 8192 intervals and a PDS for each interval was generated which were normalized such that their integral
gives the squared rms fractional variability. All the individual PDSs were then averaged to obtain the final PDS which was geometrically 
rebinned with a factor of 1.05. To model the PDS, we used a power-law model for the red noise part and Lorentzian model for QPO peaks.
 
For spectral fitting purposes, we have used 21 observations of NICER/XTI data in an energy range of $0.6-10$ keV. At first, we have done 
the fitting using the combined disk black body (DBB) and power-law (PL) models (DBB+PL). We have used the photoelectric absorption model 
`\textit{PHABS}' for spectral analysis. After the {\fontfamily{qcr}\selectfont PHABS(DBB+PL)} model fitting, we have refitted all the 
observations using the TCAF model in the same energy range. We have frozen the hydrogen column density $n_H$ to 
$0.5 \times 10^{22} ~atoms~cm^{-2}$ for both the models. A systematic error of $1\%$ has been used for the spectral fitting.

\subsection{AstroSat/LAXPC}

From the \href{https://www.tifr.res.in/~antia/laxpc.html}{LAXPC Software website} we use the publicly available code to run the extraction 
process to generate level-2 files to create source and background spectra for LAXPC10 unit (unit 1) using all anode layers. 
The `\textit{laxpcl1.f}' program is run first to process multiple orbits of level-1 data. This produces event files, light curves, spectra, and GTI 
files in both the ASCII and FITS format. Afterwards, the `\textit{backshiftv3.f}' program is run to apply background correction to the light curve 
and spectrum files. This also identifies the response files to be used for any particular observation. 
For each observation, initially we run the program `\textit{laxpcl1.f}' with a time bin of 1 sec and full range of anodes and channels. The program
produces the output file \textit{lxp1level2.gti} which is moved to the file \textit{gti.inp}. Once the \textit{gti.inp} file is prepared, we extract 
0.01 sec binned light curve by running the program `\textit{laxpcl1.f}' again with the full range of anodes and channels. Since there are data gaps in the 
light curves, we extract 0.01 sec light curves for each of the time range (listed in the \textit{gti.inp} file) separately. Then using the task
{\fontfamily{qcr}\selectfont powspec} we produce power density spectra (PDS) from the 0.01 sec time binned light curves. 

Along with spectral fitting with NICER data in the $0.6-10$ keV energy band, we have also used two observations from LAXPC data to check our 
spectral fitting in a broad range. We have used combined NICER+LAXPC data for the simultaneous observations of both the satellites/instruments 
on 2017 October 4 and 2017 October 5. We have also used {\fontfamily{qcr}\selectfont PHABS(DBB+PL)} and {\fontfamily{qcr}\selectfont PHABS*TCAF} 
models in a range of $0.6-28$ keV energy range, using a systematic error of $1\%$ and freezing the column density at $0.5 \times 10^{22} ~atoms~cm^{-2}$.

\subsection{NuSTAR}

To extract NuSTAR data, we have made use of the \textsc{NuSTARDAS (v1.4.1)} software. We first run the \textit{nupipeline} command to produce
stage-II cleaned event files. Then using \textsc{XSELECT} task, we choose region files for the source and background with the help of a circular
region of 80 $arcsec$ radius. Using those region files, we run the \textit{nuproducts} command to produce spectrum and light curve files. 

For the spectral analysis, we have used only 2 observations on 2017 October 25 (80302304004) and 2017 November 2 (80302304005) respectively.

\vspace{0.5cm}
\section{Results}

We studied detailed spectral and temporal properties of the 2017-18 outburst of the GX 339-4 using multiple satellite data. We first studied the timing 
properties of the source using MAXI/GSC, Swift/BAT, NICER/XTI, AstroSat/LAXPC, and NuSTAR data. The variation of the hard and soft band fluxes of BAT and GSC,
and their ratio as hardness ratio (HR), allowed us to get a rough idea about the spectral state and flow properties of the source (Fig. 1). NICER and 
AstroSat data were used to search for the signature of one of the most interesting temporal features i.e., QPO in the PDS (Fig. 2).
To infer more about the accretion flow dynamics, the spectral nature of the source was studied, first with the phenomenological models (Fig. 1d), and then 
with the physical model (Fig. 3 \& 4). The spectral analysis is done using NICER, AstroSat/LAXPC, and NuSTAR data respectively.
In the following sub-Sections, results obtained from our analysis are discussed in detail.

\subsection{Outburst Profile}

\begin{figure*}
\vskip 0.2cm
  \centering
    \includegraphics[angle=0,width=14cm,keepaspectratio=true]{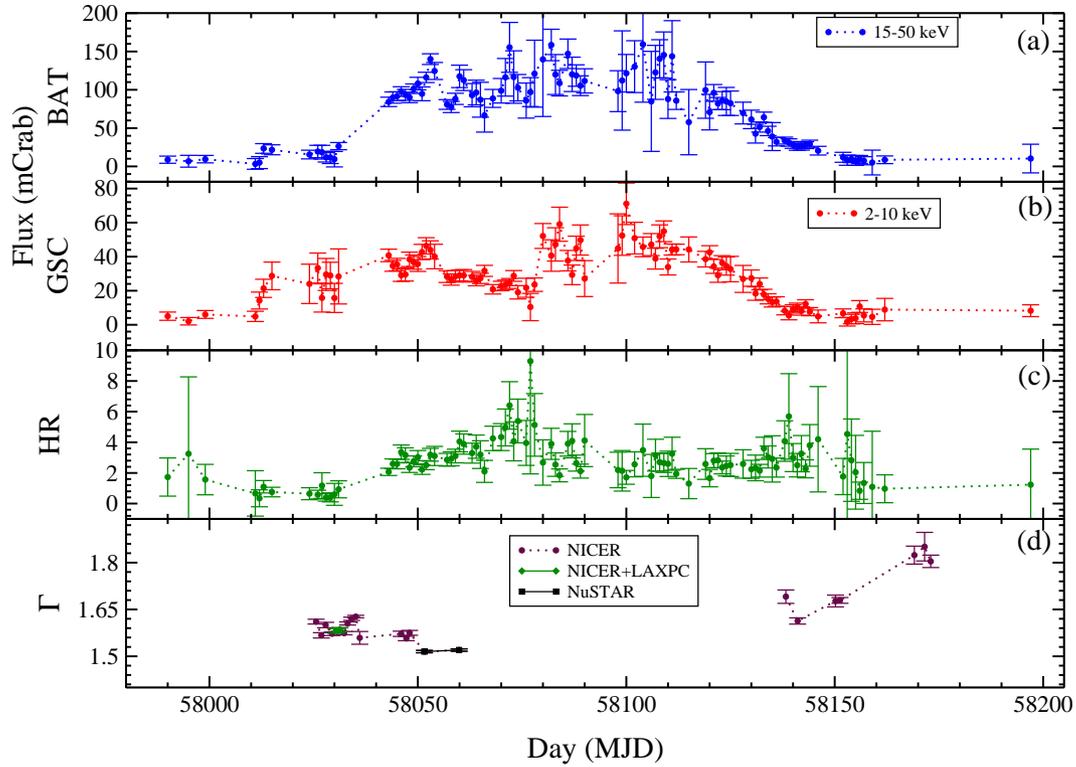}
	\caption{Variation of (a) mCrab converted $15-50$ keV Swift/BAT (blue), (b) $2-10$ keV MAXI/GSC (red) fluxes, (c) hardness ratio (HR, i.e., ratio of BAT and GSC fluxes). The variation of the power law model fitted indices ($\Gamma$s) are plotted in (d).}
\end{figure*}

The variation of the Swift/BAT (in 15-50 keV band), MAXI/GSC (in 2-10 keV band) from 2017 August 25 (MJD= 57990) to 2018 March 20 (MJD=58197) are shown
in Fig. 1(a-b). Although in the initial rising and late declining phases of the outburst, both soft and hard X-ray fluxes were roughly comparable, hard 
X-ray flux was more dominating in the middle phase of the outburst. Overall, if we look at the values of the hardness-ratios (HRs) and their variation 
(Fig. 1c), we see that during the entire period of the outburst, the source was dominated by the higher energy photons. This indicates spectral nature of 
the source might be hard. To confirm this, we did the spectral analysis.

\subsection{Detection of multiple harmonic QPO}

We used NICER and AstroSat data for searching QPOs in the PDS, generated with the $0.01$~s lightcurves. We found a prominent signature of QPOs only in one
observation of NICER. No QPOs were found in the LAXPC data. The NICER observed QPO is an interesting one as it contains one primary QPO of $\sim 0.31$~Hz
and its four harmonics. 
To fit the PDS in XSPEC, we converted the XRONOS package generated PDS data following the method of Ingram \& Done (2012).
We initially fitted the broadband noise continuum with a powerlaw model and successively added Lorentzian profiles to improve the fit.
To confirm the existence of the harmonics and their significance, we used the `{\fontfamily{qcr}\selectfont ftest}' method in XSPEC to check
how significantly the fit improved after adding one more Lorentzian profile to the previous model in each step (see, Table 1).
An acceptable fit was obtained with a model consisting of a powerlaw plus four Lorentzians with a reduced chi-square $\sim 1.53$ (Fig. 2a). 
After adding one more Lorentzian at $\sim 2.1$~Hz, the reduced chi-square becomes better $\sim 1.36$ (Fig. 2b). 
From the results of F-test, as listed in Table 1,
the addition of the first two Lorentzians to the original powerlaw model does not seem statistically necessary, but they appear to be prominent as seen in Figure 2. That is because the fitting improvement is not significant enough due to the presence of other possible peaks.
The addition of the third and fourth Lorentzian peaks then improved the fitting more significantly.
Based on the F-test, the presence of a fifth peak at about 2.1 Hz seems also statistically significant. 

\begin{figure*}
\vskip 0.2cm
\centering
\vbox{
\includegraphics[width=8.0truecm,angle=0]{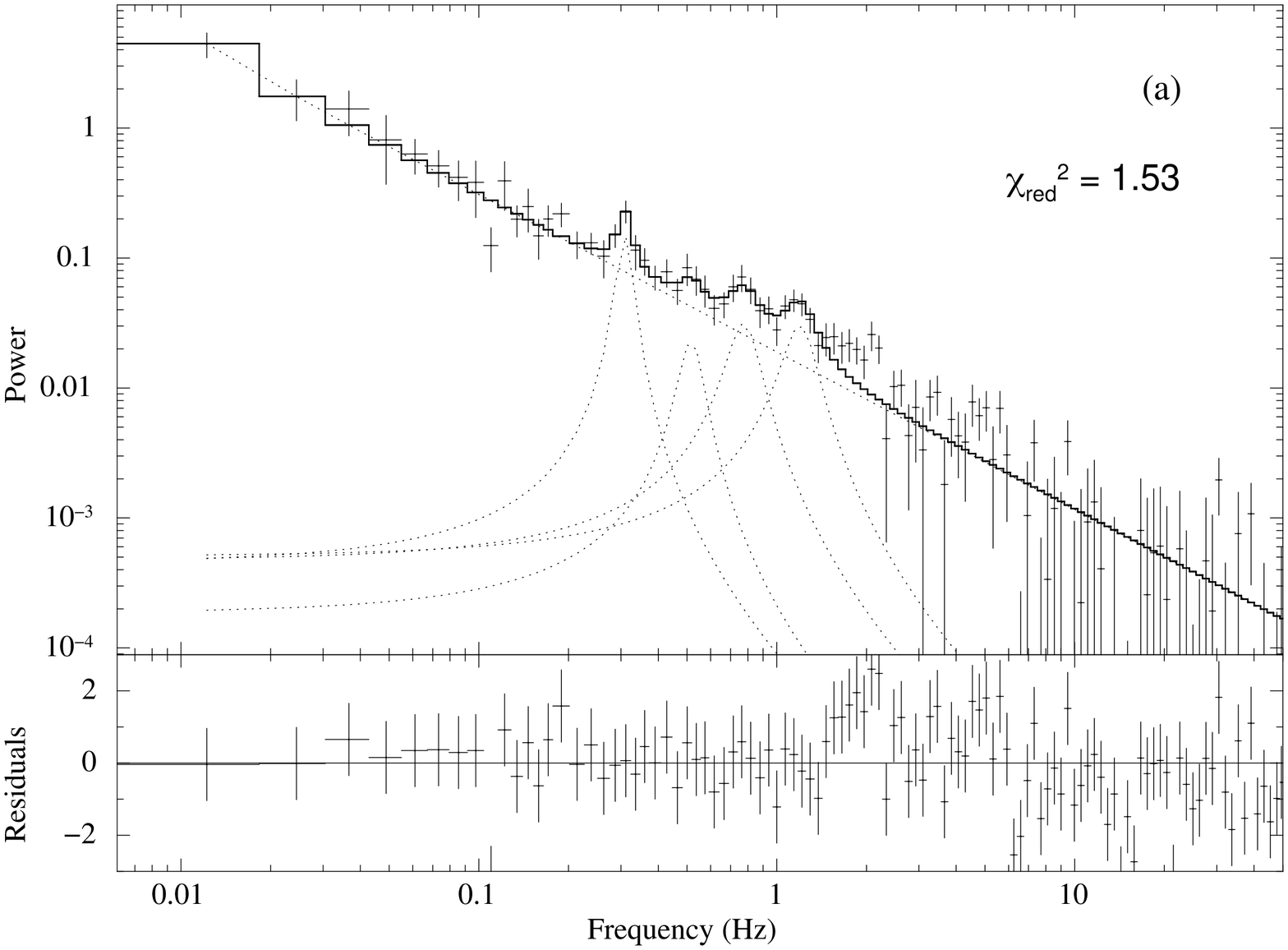}\hskip 0.5cm
\includegraphics[width=8.0truecm,angle=0]{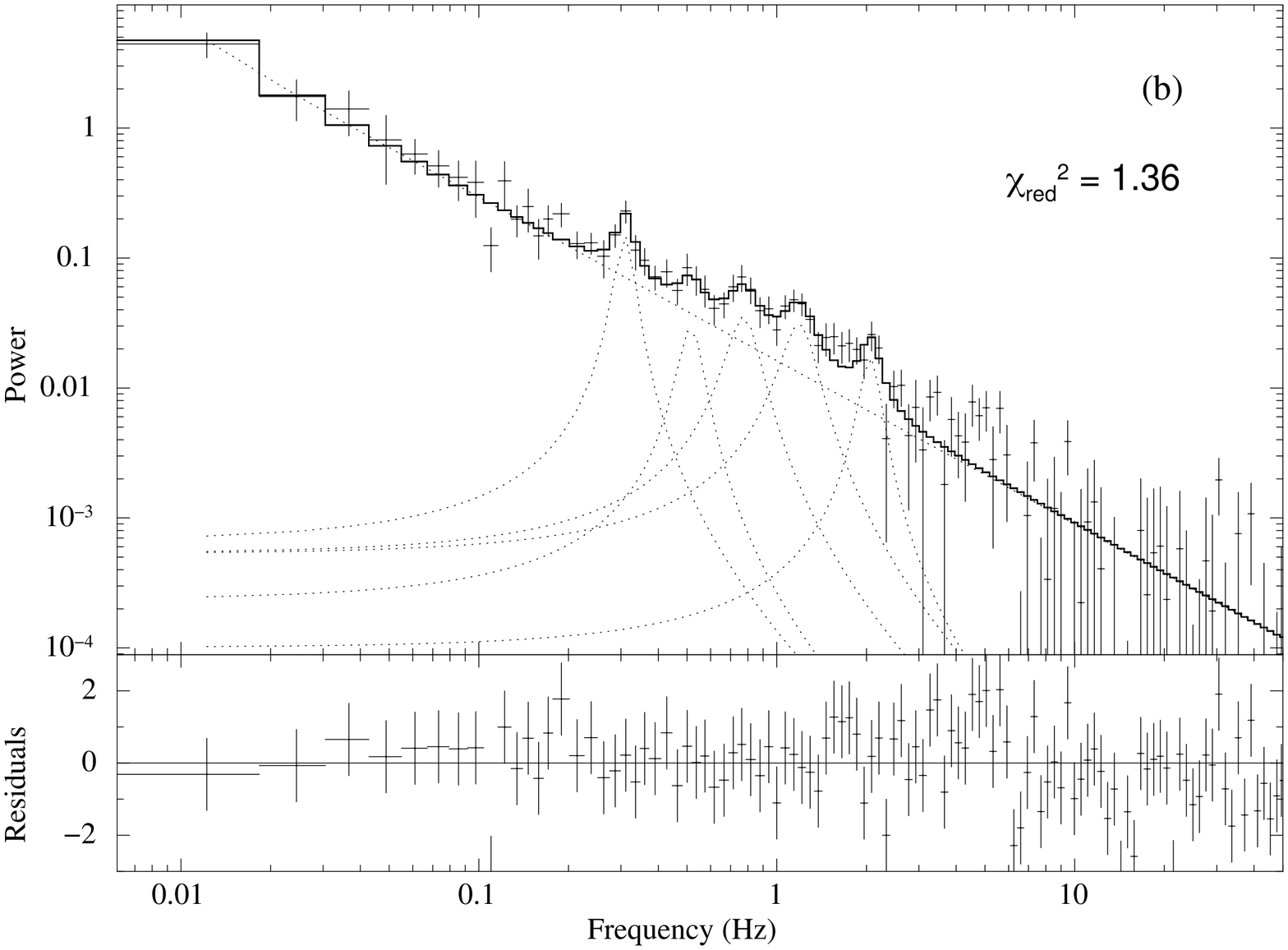}}
\vskip 0.1cm
	\caption{NICER Continuum ($0.012-50$ Hz) fitted power density spectra (PDS) of observation ID 1133010110 (UT Date: 2017-10-09 \&  MJD = 58035.10). 
	In the left panel (a) PDS is fitted with one power-law plus four Lorentzian models (for primary QPO of $0.31$~Hz and its three harmonics) and in the 
	right panel (b) one additional Lorentzian model to fit QPO of $\sim 2.06$~Hz. Model fitted reduced $\chi^2$ i.e., $\chi^2_{red}$ values are marked inset.} 
\end{figure*}

\subsection{Spectral Properties}

Although temporal data of GSC and BAT were available roughly on a daily basis, spectral data of NICER/XTI, Swift/XRT, AstroSat/LAXPC, NuSTAR were available for 
a limited observational period of the outburst. Since XTI and XRT have similar energy ranges and dates of the observation, here we used NICER data for our 
detailed spectral analysis. There were roughly one month of observations each in the both rising (from 2017 Sep. 29 to Oct. 22) and declining (from 2018 
Jan. 20 to Feb. 24) phases of the outburst. LAXPC continuously observed the source only for two days (4-5 Oct., 2017). During the entire outburst, NuSTAR 
observed the source only four times.

We first fitted spectra with the phenomenological DBB plus PL models. DBB model was required only in 6 of our studied 21 NICER observations. Both the NuSTAR 
observations were fitted with the `DBB+PL' model.
In Fig. 1(d), we show the variation of the PL model fitted photon index ($\Gamma$) values. During the entire phase of the outburst, $\Gamma$ is found to vary 
in between $1.55-1.80$. These low $\Gamma$ values indicate a harder spectral nature of the source. This is also evident from the non-requirement of the 
thermal DBB component in most of the spectra.

A sample TCAF model fitted spectrum of 2017 Oct. 5 (MJD=58031) using combined NICER and LAXPC data are shown in Fig. 3. While fitting spectra with the 
physical TCAF model, we also observed high dominance of the sub-Keplerian halo rate (${\dot m}_h$) compared to the Keplerian disk rate (${\dot m}_d$). 
The variations of the TCAF model fitted parameters are shown in Fig. 4. The rising and declining phase results are plotted in left and right panels 
respectively. Two green points in the left panels show combined NICER and LAXPC spectral analysis results while two magenta points show 
the same for two NuSTAR observations respectively. We use 5-28 keV LAXPC data of orbits 10916 
(UT: 2017-10-04, MJD=58030.39) and 10925 (UT: 2017-10-05, MJD=58031.29) are used to have a combined energy range of 1-28 keV (using NICER data of 1-8 keV) 
for these observations. We have not considered LAXPC data above 28 keV due to low signal-to-noise ratio in the data. We fitted NuSTAR data in 3-70 keV.
Model fitted spectral analysis 
results are mentioned in Table 2.

Initially, in the rising phase, higher ${\dot m}_h$ was found to be constant at $\sim 0.4$~${\dot M}_{Edd}$, then in the last three observations, it 
increased to $\sim 0.8$~${\dot M}_{Edd}$. In the declining phase, the halo rate was also observed at $\sim 0.8$~${\dot M}_{Edd}$ for the initial two 
observations, then it decreased to a lower value same as the start of the outburst, i.e., at $\sim 0.4$~${\dot M}_{Edd}$. Throughout the outburst,
${\dot m}_d$ was observed in a narrow range of lower values ($0.07-0.08$~${\dot M}_{Edd}$). 
In the initial rising phase of the outburst, the shock was found far away ($\sim 94~r_s$) from the black hole with stronger strength ($R>2.75$). 
In the late rising phase, shock moved to near location ($\sim 55~r_s$) with comparatively weak strength ($R\sim 1.6$). So, in the rising phase, 
shock moved inward as outburst progressed. In the declining phase, an opposite scenario was observed. Initially, in the declining phase, shock was found 
at $\sim 55~r_s$ with compression ratio $R\sim 1.6$. 

\begin{figure*}
\vskip 0.2cm
  \centering
    \includegraphics[angle=270,width=10cm,keepaspectratio=true]{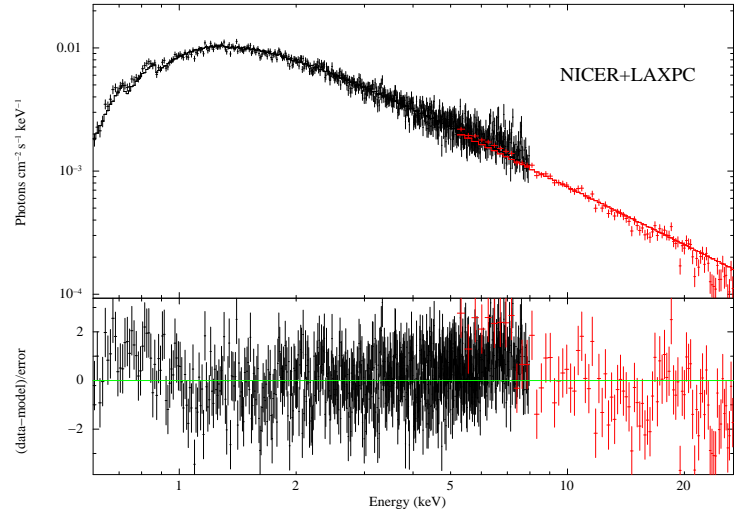}
	\caption{TCAF model fitted NICER+LAXPC spectrum in the 0.6--28 keV energy band on 2017 Oct. 5 (MJD=58031.29). Black points are for NICER data in $0.6-8$~keV 
	band and red points are for LAXPC data in $5-28$~keV band.}
\end{figure*}

\begin{figure*}
\vskip 0.2cm
  \centering
    \includegraphics[angle=0,width=14cm,keepaspectratio=true]{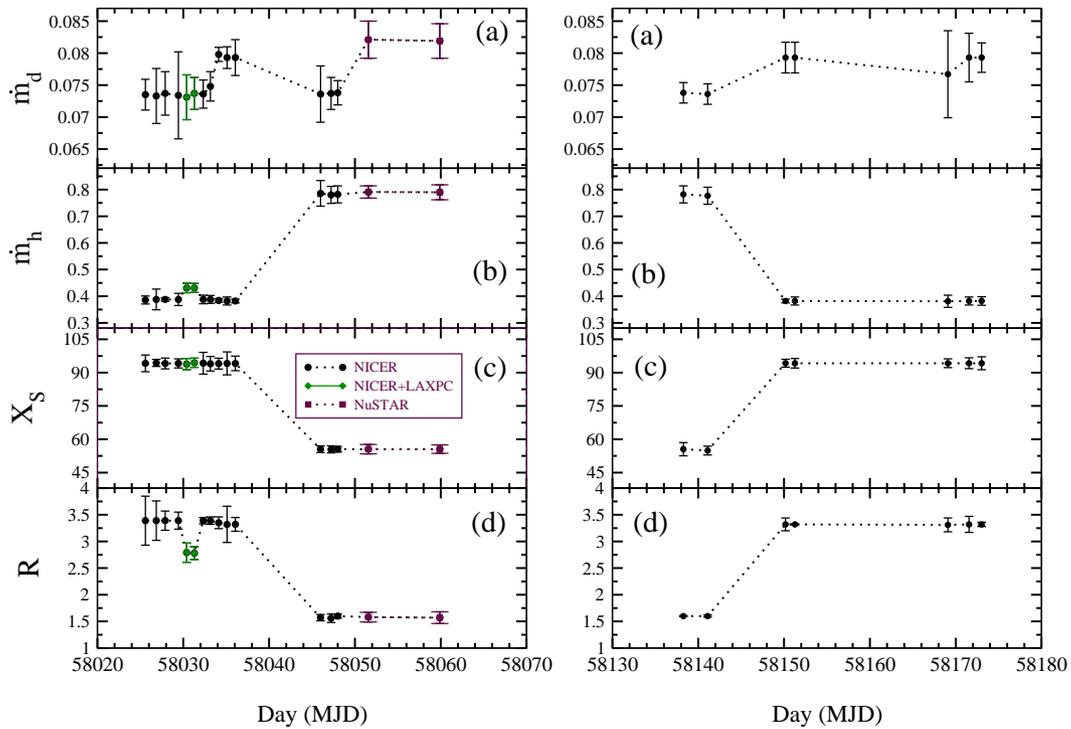}
	\caption{Variations of TCAF model fitted (a) disk rate (${\dot m}_d$), (b) halo rate (${\dot m}_h$), (c) shock location ($X_s$) and (d) compression ratio ($R$)
	  are shown. In the left and right panels parameter variations of the rising and declining phases of the outbursts are shown respectively.}
\end{figure*}

\subsection{Estimation of Mass}

Mass, which is one of the intrinsic properties of a black hole, has been included as an input parameter in the TCAF model \textit{fits} file 
(Debnath et al. 2014, 2015a). While doing a spectral fitting with the TCAF model, we may keep the mass fixed or free, depending on if it has been known from previous 
measurements or not. For this source, we have kept the mass free to estimate the mass of the BHC. We obtained mass variation in between 
$9.99-11.83$~$M_\odot$. Instrumental error plays a major role in this variation as it could be different for different observations. 
Since mass is dependent on the fourth power of the disk temperature, it is highly sensitive to any small error in the measurement of disk 
temperature. Thus, this may not be the actual variation of mass rather the variation due to instrumental error during the period of our analysis. 
Keeping this fact in mind, we have found the most probable mass of the source as $10.76^{+0.77}_{-1.07}~M_\odot$ during this outburst.
Here, $10.76~M_\odot$ is the average value of the model fitted masses.

\section{Discussion and Concluding Remarks}

GX 339-4 is a recurring Galactic transient BHC. This is a well studied black hole binary by several groups of authors in a wide band of energies from radio 
to $\gamma$-rays. Recently, Bhowmick et al. (2021) studied outburst profiles of the source to know about the triggering mechanism of the outbursts. 
A good linear relation between the outbursts and the quiescence phases is found from their analysis. A similar long-term variability of the source was also 
studied by Aneesha et al. (2019). In the past two and half decades of the post RXTE era, GX 339-4 showed outbursts of both type-I \& II. According to
Debnath et al. (2017), in type-I or complete outbursts, all four commonly identified spectral states (HS, HIMS, SIMS, SS) are found to form hysteresis 
loop, while in the type-II or harder spectral states, we do not see any soft spectral state. Sometimes even intermediate states are found to be missing 
in the type-II outbursts. Our present studied 2017-18 outburst is a `failed' in nature as only HS was found during the entire outburst.

Both the spectral and temporal nature of the source during its 2017-18 outburst was studied using archival data of MAXI, Swift, NICER, AstroSat, and NuSTAR
instruments. The variation of hard and soft band photo fluxes using Swift/BAT and MAXI/GSC data and their ratio as HR, we concluded that during the 
entire phase of the outburst source was in the harder spectral states (Fig. 1). To understand more detail about the accretion flow dynamics of the 
source during the outburst, we did spectral analysis. Due to the unavailability of the daily observations of Swift/XRT, NICER/XTI, AstroSat/LAXPC, and NuSTAR
detailed timing and spectral studies were done for limited period of the rising and declining phases of the outburst.

Low-frequency QPOs are one of the characteristic signatures of the transient BHCs. We used NICER and AstroSat/LAXPC data for searching QPOs in the 
fast-Fourier transformed PDS. During the entire outburst, in only one observation (ID: 1133010110) of NICER on 2017 Oct. 09 (MJD = 58035.10) we found 
a prominent QPO of $0.31$~Hz. Most interestingly, this primary QPO showed four of its harmonics 
(see Fig. 2). The significance of the QPO and its harmonics are tested using {\fontfamily{qcr}\selectfont ftest} method (see Table 1).

The spectral analysis was done with two types of models: i) phenomenological DBB plus PL models, and ii) physical TCAF model. The phenomenological 
model fit provides us a rough idea about the variation of the thermal and non-thermal component of the photon flux. But the detailed physics of the 
generation of such photons is missing. The physical model fit gives us more information about the flow dynamics of the source. Since the source 
was highly dominated with the non-thermal fluxes, PL model was sufficient to fit the spectra, except for 6 observations of the middle phase of 
the outburst. If there were more observations in this phase of the outburst, we might require the DBB model component in more observations. In Table 2, 
spectral analysis results are mentioned.

The variation of the PL photon index ($\Gamma$) in the range of $1.55-1.80$ (Fig. 1d) signifies a harder spectral nature of the source during our 
analysis period. While fitted spectra with the TCAF model, we also noticed high dominance of the sub-Keplerian halo rate. Low values of the Keplerian 
disk rate (compared to the halo rate) confirm the non-requirement of the DBB model in most of the observations while spectra were fitted with the 
phenomenological models. These higher halo rates confirm the spectral nature of the source as HS. In the early rising and late declining phase of 
the outburst, a stronger shock was found at a comparatively large distance from the black hole. In the middle phase, the shock was found to be weaker 
and nearer.

We also estimated the probable mass of the black hole ($M_{BH}$) from our spectral analysis with the TCAF model. Since in the TCAF model, $M_{BH}$ is 
an important input parameter. So, if it is not well known and kept as free while fitting spectra, we get one best-fitted value of the $M_{BH}$. 
The variation of the $M_{BH}$ during the outburst does not mean the real variation of the $M_{BH}$ during our analysis period. The $M_{BH}$ is highly 
sensitive to the disk temperature (T). If there is an error in the measurement of $T$, it contributes a large variation in $M_{BH}$ since $M_{BH} \sim T^4$. 
We observed the variation of the $M_{BH}$ in a range of $9.99-11.83$~$M_\odot$ or $10.76^{+0.77}_{-1.07}~M_\odot$. Here, $10.76~M_\odot$ is the 
average of the model fitted $M_{BH}$. This estimated mass of GX 339-4 is consistent with the earlier findings of Parker et al. (2016) and 
Sreehari et al. (2019).

\section{Summary}

We have studied the timing and spectral properties of the $2017-18$ outburst of GX 339-4 using publicly available archival data from MAXI/GSC, Swift/BAT, 
NICER/XTI, AstroSat/LAXPC, and NuSTAR/FPMA satellite instruments. Using $0.01~sec$ time binned light curves data from XTI ($1-10~ keV$) and 
LAXPC ($3-80~keV$) instruments, we have searched for quasi periodic oscillations during the entire phase of the outburst. We have also performed spectral 
analysis using only NICER ($0.6-10~keV$) data for 19 observations, combined NICER+LAXPC ($0.6-28~keV$) data for 2 observations and NuSTAR ($3-70~keV$) 
for 2 more observations. Due to the absence of data in a period of almost $2.5$~months in the middle phase of the outburst, we were unable to define any 
transition between the spectral states properly. 
However, based on the variations of HR and dominance of hard flux over soft flux, we have characterized the state of the source as hard. 
It did not make any transition into the defined softer states during our studied period of the outburst. While fitting the spectra, we have 
also estimated the mass of this source during this outburst. Based on our temporal and spectral analysis, we conclude that:

(i) The $2017-18$ is a failed or harder type of outburst, which does not make any transition into the softer states.

(ii) The source has a mass between $9.99-11.83~M_\odot$, which is in agreement with previous findings.

(iii) We have found QPO only in one day (2017 October 9; MJD=58035) of the entire outburst. The observed QPO is an interesting one as 
it showed multiple harmonics.

\section{Data Availability}

We have used archival data of \href{http://maxi.riken.jp/top/lc.html}{MAXI/GSC}, \href{https://swift.gsfc.nasa.gov/results/transients/}{Swift/BAT},
\href{https://heasarc.gsfc.nasa.gov/db-perl/W3Browse/w3browse.pl}{NICER}, 
\href{https://heasarc.gsfc.nasa.gov/db-perl/W3Browse/w3browse.pl}{NuSTAR}
and \href{https://astrobrowse.issdc.gov.in/astro_archive/archive/Home.jsp}{AstroSat/LAXPC} 
for this work.

\section*{Acknowledgements}
This research has made use of data and/or software provided by the High Energy Astrophysics Science Archive Research Center (HEASARC),
which is a service of the Astrophysics Science Division at NASA/GSFC. 
This work made use of Swift/BAT data supplied by the UK Swift Science Data Centre at the University of Leicester; MAXI/GSC data provided 
by RIKEN, JAXA, and the MAXI team; NICER data archived by NASA/GSFC; AstroSat/LAXPC data obtained from the data archive of Indian Space 
Science Data Centre (ISSDC). We acknowledge the strong support from Indian Space Research Organization (ISRO) for the successful 
realization and operation of AstroSat mission. The authors also acknowledge the AstroSat team for the distribution. The LaxpcSoft 
software is used for the analysis. D.D. acknowledges support from DST/GITA sponsored India-Taiwan collaborative project (GITA/DST/TWN/P-76/2017). 
Research of D.D. is supported in part by the Higher Education Dept. of the Govt. of West Bengal, India.
K.C. acknowledges support from DST/INSPIRE (IF170233) fellowship. 
S.N. and D.D. acknowledge partial support from ISRO sponsored RESPOND project (ISRO/RES/2/418/17-18) fund. 
R.B. acknowledges support from CSIR-UGC fellowship (June-2018, 527223).


\clearpage

\begin{table*}[h]
\small
 \addtolength{\tabcolsep}{-1.5pt}
 \centering
 \caption{QPO Properties}
 \label{tab:table1}
        \resizebox{1 \textwidth}{!}{
        \begin{tabular}{cccccccccc}
\hline
Model  &     PL Index       &       PL Norm.       &    $\nu_{QPO}$     &   $\Delta\nu$    &        Q                & rms \%   &${\chi}^2$/dof & F-stat & Prob. \\
       &     ($\Gamma$)     &                      &       (Hz)           &      (Hz)          &   ($\nu/\Delta\nu$)     &          &               &        &      \\
  (1)  &        (2)         &        (3)           &       (4)            &      (5)           &         (6)             &   (7)    & (8)           &   (9)  & (10) \\
\hline
\hline
PL     & $ 1.16^{\pm0.02} $ & $ 0.026^{\pm0.001} $ &                      &                    &                         &          & 218/107 &        &      \\
\hline
PL+LOR & $ 1.16^{\pm0.02} $ & $ 0.025^{\pm0.001} $ & $ 0.308^{\pm0.008} $ & $ 0.02^{\pm0.01} $ &        15.4             &   1.37   & 206/104 &  2.02  & 0.12  \\
\hline
PL+LOR & $ 1.16^{\pm0.03} $ & $ 0.024^{\pm0.001} $ & $ 0.308^{\pm0.008} $ & $ 0.02^{\pm0.01} $ &        15.4             &   1.37   & 203/101 &  0.50  & 0.68 \\
+LOR   &                    &                      & $ 0.533^{\pm0.042} $ & $ 0.11^{\pm0.03} $ &         4.8             &   2.63   &         & (1.24) & (0.29)       \\ 
\hline
PL+LOR & $ 1.18^{\pm0.03} $ & $ 0.022^{\pm0.001} $ & $ 0.309^{\pm0.007} $ & $ 0.02^{\pm0.01} $ &        15.4             &   1.48   & 181/ 98 &  3.97  &0.010\\
+LOR   &                    &                      & $ 0.518^{\pm0.048} $ & $ 0.09^{\pm0.03} $ &         5.8             &   1.68   &         & (2.23)  & (0.026)      \\
+LOR   &                    &                      & $ 0.803^{\pm0.030} $ & $ 0.21^{\pm0.08} $ &         3.8             &   6.29   &         &        &       \\
\hline
PL+LOR & $ 1.21^{\pm0.03} $ & $ 0.019^{\pm0.002} $ & $ 0.309^{\pm0.006} $ & $ 0.03^{\pm0.01} $ &        10.3             &   1.94   & 146/ 95 &  7.59  &0.00013\\
+LOR   &                    &                      & $ 0.516^{\pm0.034} $ & $ 0.09^{\pm0.05} $ &         5.7             &   2.06   &         &  (3.90)      & (0.000073)     \\ 
+LOR   &                    &                      & $ 0.775^{\pm0.035} $ & $ 0.19^{\pm0.07} $ &         4.1             &   5.18   &         &        &      \\
+LOR   &                    &                      & $ 1.195^{\pm0.037} $ & $ 0.31^{\pm0.03} $ &         3.8             &   8.26   &         &        &       \\
\hline
PL+LOR & $ 1.26^{\pm0.04} $ & $ 0.016^{\pm0.002} $ & $ 0.310^{\pm0.006} $ & $ 0.04^{\pm0.01} $ &         7.8             &   2.51   & 126/ 92 &  4.87  &0.0035\\
+LOR   &                    &                      & $ 0.515^{\pm0.026} $ & $ 0.09^{\pm0.02} $ &         5.7             &   2.38   &         & (4.48)       & ($<10^{-5}$)     \\ 
+LOR   &                    &                      & $ 0.773^{\pm0.031} $ & $ 0.19^{\pm0.04} $ &         4.1             &   5.46   &         &        &      \\
+LOR   &                    &                      & $ 1.188^{\pm0.036} $ & $ 0.31^{\pm0.03} $ &         3.8             &   8.55   &         &        &      \\
+LOR   &                    &                      & $ 2.061^{\pm0.044} $ & $ 0.21^{\pm0.08} $ &         9.8             &   5.45   &         &        &      \\
\hline
\hline
        \end{tabular}}
\vskip 0.2cm
\leftline{In column (1), PL and LOR mean Powerlaw and Lorentzian models respectively.}
\leftline{In columns (4) and (5), $\nu_{QPO}$ is the QPO frequency and $\Delta\nu$ is the full width at half maximum (FWHM).}
\leftline{In columns (9) and (10), F-stat and Prob. mean $ftest$ statistics and random probabilities of the present model with}
\leftline{the previous model. Those in parentheses are of the present model with the PL-only model.}
\end{table*}

\begin{table}
\small 
 \addtolength{\tabcolsep}{-1.0pt}
 \centering
 \caption{DBB+PL and TCAF model fitted parameters}
 \label{tab:table2}
\resizebox{1 \textwidth}{!}{
 \begin{tabular}{|ccc|ccc|cccccc|}
 \hline
Obs. ID$^{[1]}$  & UT$^{[1]}$  &  MJD$^{[1]}$    &     T$_{in}$$^{[2]}$        & $\Gamma$$^{[2]}$       & ${\chi}^2$/dof$^{[4]}$ &  ${\dot m}_d$$^{[3]}$    &  ${\dot m}_h$$^{[3]}$  &  $X_s$$^{[3]}$      &     $R$$^{[3]}$      &  $M_{BH}$$^{[3]}$     &  ${\chi}^2$/dof$^{[4]}$      \\ 
    (1)          &    (2)      &     (3)         &           (4)               &       (5)              &          (6)           &          (7)             &         (8)            &         (9)         &        (10)          &       (11)            &           (12)               \\
\hline

    N01          & 2017-09-29  &  58025.58     &   $      -     $            &  $1.61^{\pm 7.7e-3} $  &  $    561/608  $    &  $0.0735^{\pm 0.0024} $  &  $0.386^{\pm 0.015} $  &  $94.1^{\pm 3.7} $  &  $3.39^{\pm 0.46} $  &  $11.48^{\pm 0.99} $  &    $  735/604        $    \\
    N02          & 2017-09-30  &  58026.85     &   $      -     $            &  $1.56^{\pm 8.4e-3} $  &  $    590/608  $    &  $0.0733^{\pm 0.0043} $  &  $0.388^{\pm 0.039} $  &  $94.3^{\pm 1.6} $  &  $3.39^{\pm 0.37} $  &  $11.42^{\pm 1.06} $  &    $  724/579        $    \\ 
    N03          & 2017-10-01  &  58027.88     &   $      -     $            &  $1.60^{\pm 8.6e-3} $  &  $    701/608  $    &  $0.0737^{\pm 0.0034} $  &  $0.388^{\pm 0.007} $  &  $94.1^{\pm 2.4} $  &  $3.39^{\pm 0.18} $  &  $11.45^{\pm 1.56} $  &    $ 1008/736        $    \\ 
    N04          & 2017-10-03  &  58029.43     &   $      -     $            &  $1.57^{\pm 5.6e-3} $  &  $    712/698  $    &  $0.0734^{\pm 0.0068} $  &  $0.388^{\pm 0.023} $  &  $94.1^{\pm 2.0} $  &  $3.39^{\pm 0.16} $  &  $11.47^{\pm 0.68} $  &    $  754/664        $    \\ 
    N05$^*$      & 2017-10-04  &  58030.39     &   $      -     $            &  $1.58^{\pm 7.0e-3} $  &  $    692/616  $    &  $0.0731^{\pm 0.0035} $  &  $0.431^{\pm 0.018} $  &  $93.8^{\pm 2.5} $  &  $2.79^{\pm 0.18} $  &  $11.79^{\pm 0.36} $  &    $  898/699        $    \\ 
    N06$^*$      & 2017-10-05  &  58031.29     &   $      -     $            &  $1.58^{\pm 6.3e-3} $  &  $    714/655  $    &  $0.0737^{\pm 0.0025} $  &  $0.431^{\pm 0.017} $  &  $94.4^{\pm 2.0} $  &  $2.78^{\pm 0.12} $  &  $11.83^{\pm 0.28} $  &    $ 1061/739        $    \\ 
    N07          & 2017-10-06  &  58032.32     &   $      -     $            &  $1.57^{\pm 5.3e-3} $  &  $    743/715  $    &  $0.0736^{\pm 0.0022} $  &  $0.388^{\pm 0.016} $  &  $94.2^{\pm 4.9} $  &  $3.39^{\pm 0.06} $  &  $11.50^{\pm 0.77} $  &    $  864/711        $    \\ 
    N08          & 2017-10-07  &  58033.15     &   $      -     $            &  $1.60^{\pm 4.9e-3} $  &  $    798/739  $    &  $0.0748^{\pm 0.0023} $  &  $0.388^{\pm 0.015} $  &  $94.0^{\pm 3.2} $  &  $3.39^{\pm 0.07} $  &  $11.47^{\pm 0.27} $  &    $  832/694        $    \\ 
    N09          & 2017-10-08  &  58034.13     &   $      -     $            &  $1.62^{\pm 5.1E-3} $  &  $    842/705  $    &  $0.0798^{\pm 0.0011} $  &  $0.384^{\pm 0.007} $  &  $94.0^{\pm 2.4} $  &  $3.35^{\pm 0.11} $  &  $11.56^{\pm 0.30} $  &    $  856/701        $    \\ 
    N10          & 2017-10-09  &  58035.09     &   $      -     $            &  $1.62^{\pm 4.0E-3} $  &  $    956/766  $    &  $0.0793^{\pm 0.0017} $  &  $0.382^{\pm 0.015} $  &  $94.1^{\pm 5.1} $  &  $3.32^{\pm 0.34} $  &  $11.59^{\pm 1.24} $  &    $  934/762        $    \\ 
    N11          & 2017-10-10  &  58036.06     &   $0.256^{\pm 3.3e-2  } $  &  $1.55^{\pm 2.0e-2} $  &  $    570/569  $    &  $0.0793^{\pm 0.0028} $  &  $0.382^{\pm 0.007} $  &  $94.1^{\pm 3.2} $  &  $3.32^{\pm 0.13} $  &  $11.59^{\pm 1.78} $  &    $  597/567        $    \\ 
    N12          & 2017-10-20  &  58046.00     &   $0.235^{\pm 5.7e-3  } $  &  $1.57^{\pm 8.3e-3} $  &  $    879/791  $    &  $0.0736^{\pm 0.0044} $  &  $0.786^{\pm 0.048} $  &  $55.5^{\pm 1.4} $  &  $1.57^{\pm 0.06} $  &  $10.02^{\pm 0.75} $  &    $  905/749        $    \\ 
    N13          & 2017-10-21  &  58047.22     &   $0.239^{\pm 4.4e-3  } $  &  $1.55^{\pm 7.1e-3} $  &  $    901/838  $    &  $0.0737^{\pm 0.0025} $  &  $0.780^{\pm 0.032} $  &  $55.4^{\pm 1.6} $  &  $1.56^{\pm 0.08} $  &  $9.99 ^{\pm 0.34} $  &    $ 1038/796        $    \\ 
    N14          & 2017-10-22  &  58048.00     &   $0.224^{\pm 5.1e-3  } $  &  $1.57^{\pm 7.6e-3} $  &  $    934/831  $    &  $0.0738^{\pm 0.0019} $  &  $0.782^{\pm 0.032} $  &  $55.5^{\pm 1.2} $  &  $1.60^{\pm 0.04} $  &  $10.00^{\pm 0.27} $  &    $ 1255/789        $    \\ 
    NS04         & 2017-10-25  &  58051.57     &   $0.223^{\pm 5.1e-3  } $  &  $1.52^{\pm 3.7e-3} $  &  $   1225/1026 $    &  $0.0821^{\pm 0.0029} $  &  $0.791^{\pm 0.023}$   &  $55.6^{\pm 2.1} $  &          $1.58^{\pm 0.09}$   &  $11.52^{\pm 0.51} $  &    $ 1259/1024       $    \\
    NS05         & 2017-11-02  &  58059.89     &   $0.217^{\pm 5.1e-3  } $  &  $1.52^{\pm 3.8e-3} $  &  $   1058/1000 $    &  $0.0819^{\pm 0.0027} $  &  $0.790^{\pm 0.028}$   &  $55.6^{\pm 1.9} $  &          $1.57^{\pm 0.11}$   &  $11.52^{\pm 0.32} $  &    $ 1084/998        $    \\ 
    N21          & 2018-01-20  &  58138.27     &   $0.180^{\pm 2.2e-2  } $  &  $1.69^{\pm 2.1e-2} $  &  $    547/464  $    &  $0.0738^{\pm 0.0016} $  &  $0.782^{\pm 0.032} $  &  $55.5^{\pm 2.9} $  &  $1.60^{\pm 0.01} $  &  $10.00^{\pm 0.76} $  &    $  483/422        $    \\ 
    N24          & 2018-01-23  &  58141.10     &   $0.172^{\pm 7.9e-3  } $  &  $1.61^{\pm 1.0e-2} $  &  $    689/681  $    &  $0.0736^{\pm 0.0016} $  &  $0.777^{\pm 0.032} $  &  $54.9^{\pm 1.9} $  &  $1.60^{\pm 0.02} $  &  $11.03 ^{\pm 0.24} $  &    $  680/639        $    \\ 
    N28          & 2018-02-01  &  58150.17     &   $      -     $            &  $1.67^{\pm 1.9e-2} $  &  $    339/322  $    &  $0.0793^{\pm 0.0024} $  &  $0.382^{\pm 0.007} $  &  $94.2^{\pm 1.7} $  &  $3.32^{\pm 0.12} $  &  $11.59^{\pm 0.46} $  &    $  375/318        $    \\ 
    N29          & 2018-02-02  &  58151.26     &   $      -     $            &  $1.67^{\pm 9.1e-3} $  &  $    605/541  $    &  $0.0793^{\pm 0.0024} $  &  $0.382^{\pm 0.015} $  &  $94.2^{\pm 2.1} $  &  $3.32^{\pm 0.01} $  &  $11.59^{\pm 0.39} $  &    $  698/537        $    \\ 
    N33          & 2018-02-20  &  58169.09     &   $      -     $            &  $1.82^{\pm 2.8e-2} $  &  $    261/276  $    &  $0.0767^{\pm 0.0068} $  &  $0.381^{\pm 0.023} $  &  $94.2^{\pm 1.9} $  &  $3.31^{\pm 0.13} $  &  $11.60^{\pm 0.75} $  &    $  393/264        $    \\ 
    N35          & 2018-02-22  &  58171.54     &   $      -     $            &  $1.85^{\pm 4.5e-2} $  &  $     90/104  $    &  $0.0793^{\pm 0.0038} $  &  $0.382^{\pm 0.015} $  &  $94.2^{\pm 2.4} $  &  $3.32^{\pm 0.15} $  &  $11.59^{\pm 0.37} $  &    $  139/100        $    \\ 
    N36          & 2018-02-24  &  58173.02     &   $      -     $            &  $1.80^{\pm 1.9e-2} $  &  $    359/335  $    &  $0.0793^{\pm 0.0023} $  &  $0.382^{\pm 0.016} $  &  $94.2^{\pm 2.9} $  &  $3.32^{\pm 0.04} $  &  $11.59^{\pm 0.24} $  &    $  526/423        $    \\ 

\hline
\end{tabular}
}
\vskip 0.2cm
\scriptsize
 \noindent{
\leftline{$^{[1]}$ Column 1, 2 and 3 represent the Obs Id., date and MJD respectively. `N' and `NS' in column 1 mark the initial part for the NICER and NuSTAR Obs.} \leftline{Ids respectively, where 
                   `N'=11330101 and `NS'=803023040. } 
\leftline{$^{[2]}$ Column 4 and 5 represent the DBB+PL fitted innerdisk temperature ($T_{in}$) and photon index of power-law ($\Gamma$) respectively.} 
\leftline{$^{[3]}$ Column 7, 8, 9, 10 and 11 represent the TCAF model fitted parameters: disk rate (${\dot m}_d$ in Eddington rate (${\dot M}_{Edd}$), halo rate (${\dot m}_h$ in ${\dot M}_{Edd}$),} 
\leftline{shock location ($X_s$ in Schwarzschild  radius $r_s$), compression ratio ($R$) and mass of the black hole ($M_{BH}$ in solar mass $M_{\odot}$) respectively.} 
\leftline{$^{[4]}$ DBB+PL and TCAF models fitted ${\chi}^2_{red}$ values are mentioned in columns 6 and 12 respectively as ${\chi}^2/dof$, where `dof' 
                   represents degrees of freedom.}
\leftline{  `$^*$' sign in Column 1 represents the obs. Ids for which the NICER+LAXPC combined fitting have been done. For both observations, we have taken } 
\leftline{LAXPC data with obs. Id = A04-109T01-9000001578. N05 and N06 correspond to AstroSat orbits of 10916 and 10925 respectively.}
}
\end{table}

\end{document}